\documentclass[a4paper,12pt]{article}
\usepackage{epsfig} 

\begin{document} 

{\thispagestyle{empty}
\setcounter{page}{1}

\title{Fragmentation of Colliding Discs}
\author{Ferenc~Kun$^1$ and Hans~J.~Herrmann$^2$} 
 
\maketitle
\centerline{${}^1$Laboratoire de Physique et M\'ecanique des Milieux 
            H\'et\'erog\`enes, C.N.R.S., U.R.A. 857} 
\centerline{\'Ecole Sup\'erieure de Physique et Chimie
  Industrielle}
\centerline{10 rue Vauquelin, 75231 Paris, Cedex 05, France}
\vspace*{0.5cm}
\centerline{${}^2$ICA 1, University of Stuttgart, Pfaffenwaldring 27, 
70569 Stuttgart, Germany} 

\begin{abstract}
We study the phenomena associated
with the low-velocity impact of two solid discs of equal size using a
cell model of brittle solids. The fragment ejection exhibits a jet-like
structure the direction of which depends on the impact parameter.
We obtain the velocity and the mass
distribution of the debris. Varying the radius and the initial
velocity of the colliding particles, the velocity components of the
fragments show anomalous
scaling. The mass distribution follows a power law in the region of
intermediate masses. 
\end{abstract}

\section{Introduction}
\vspace*{-0.5pt}
\noindent
Fragmentation covers a wide diversity of physical phenomena. 
Recently, the fragmentation of granular solids has attracted
considerable scientific and industrial interest. The length scales
involved in this process range from the collisional evolution of
asteroids\cite{exp_tur} to the degradation of materials comprising
small agglomerates employed in industrial processes\cite{thornton}.
On the intermediate scale there are several geophysical examples
concerning to the usage of explosives in mining and oil shale
industry, fragments from weathering, coal heaps rock fragments from
chemical and nuclear explosions\cite{exp_tur}.   
Most of the measured fragment size distributions exhibit power law
behavior with exponents between 1.9 and 2.6 concentrating around
2.4\cite{exp_tur}. Power law behavior of small fragment masses seems to be
a common characteristic of brittle fracture. 

Comprehensive laboratory experiments were carried out applying projectile
col\-li\-si\-on\cite{exp_colli1,exp_colli2,exp_colli3} and free fall impact
with a mas\-sive pla\-te\cite{exp_glass,exp_bohr,exp_meibom}. 
The resulting fragment size distributions show universal power law
behavior. The scaling exponents depend on the overall morphology of
the objects but are independent of the type of the materials. 

Beside the size distribution of the debris, there is also particular
interest in the energy required to achieve a certain size
reduction. Collision experiments\cite{exp_colli2} revealed that the
mass of the largest fragment normalized by the total mass shows power
law behavior as a function of the specific energy (imparted energy
normalized by the total mass). The exponents are between 0.6 and 1.5
depending on the geometry of the system.    

On microscopic scale the fragmentation of atomic nuclei is intensively
in\-ves\-ti\-ga\-ted\cite{frag_book}. In the experiments concerning the
multifragmentation of gold projectiles, a power law charge distribution of
the fragments was found independent of the target type\cite{botet}.

 Several theoretical approaches were proposed to describe
fragmentation. In stochastic models\cite{theo_glass,zhang,naim,gonzalo} power
law, exponential and log-normal distributions were obtained
depending on the dimensionality of the object and the detailed
breaking mechanism.

Discrete stochastic processes have also been studied as models for
fragmentation using cellular automata. In Ref. ~15  two- and
three - dimensional cellular automata were proposed to model power
law distributions in shear experiments on a layer of uniformly sized
fragments. 
 
The mean - field approach describes the time evolution of the
concentration $c(x,t)$ of fragments having mass $x$ through a linear
integro - differential equation\cite{redner}:
\begin{eqnarray}
  \label{mean}
  \frac{\partial c(x,t)}{\partial t} = -a(x) c(x,t) + 
\int_x^{\infty} c(y,t)a(y)f(x|y)dy
\end{eqnarray}
where $a(x)$ is the overall rate at which $x$ breaks in a time
interval $dt$, while $f(x|y)$ is the relative rate at which $x$ is
produced from the break-up of $y$. With some further assumptions on
$f(x|y)$ exact results can be obtained but in physically interesting
situations the solution is
very difficult\cite{redner}.    

Three - dimensional impact fracture processes of random materials were
modeled based on a competitive growth of cracks\cite{soc}. 
A universal power law
fragment mass distribution was found consistent with 
self - organized criticality with an 
exponent of $\frac{5}{3}$. 
 
A two dimensional 
dynamic simulation of solid fracture was performed 
using a cellular model material\cite{potapov1,potapov2}. 
The compressive failure of a
rectangular sample, the four - point shear failure of a beam and the
impact of particles with a plate and with other particles were
studied. 
 
Recently, we have established\cite{ferenc} a 
two - dimensional model
for a  deformable, breakable, granular solid by connecting
unbreakable, undeformable  elements with elastic beams, similar as in
Refs. ~18,19. 
The contacts between the
particles can be
broken according to a physical breaking rule, which takes into account
the stretching and bending of the connections\cite{ferenc}.

In this paper we apply the model to study the phenomena associated
with low-velocity impact of two solid discs
of the same size.
An advantage of our model with respect to most other fragmentation
models is that we can follow the trajectory of each fragment, which is
often of big practical importance, and that we know how much energy
each fragment carries away.  
Varying the impact parameter, the size of the colliding particles
and the initial velocities, we are mainly interested in the spatial
distribution of the fragments, in the distribution of
the fragment velocities and the fragment size. 
This is important to get deeper insight into
the collisional evolution of asteroids, small planets and planetary
rings. 
One particular interest of this experiment is that it can be
considered as a classical analog of the
deeply inelastic scattering of heavy nuclei\cite{frag_book}.   
The characteristic quantities providing  collective description of
the fragmenting system (e.g. the fragment mass distribution),
are scale invariant.
This enables us to make also some
comparison with nuclear fragmentation experiments. 

\section{Description of the simulation}

Here we give a brief overview of the basic ideas of our model and the
simulation technique used. For
details see Ref.~20.

In order to study fragmentation of granular solids we perform Molecular
Dynamics (MD) simulations in two dimensions. This method calculates
the motion of particles by solving Newton's equations. In our
simulation this is done using a Predictor-Corrector scheme.
The construction of our model of a deformable, breakable, granular
solid is performed in three major steps, namely, the implementation of
the granular structure of the solid, the introduction of the elastic
behavior by the cell repulsion and the beam model, and finally the
breaking of the solid.

 In order to take into account the complex structure of the granular
solid we use arbitrarily shaped convex polygons.
To get an initial configuration of these polygons we make a special Voronoi
tessellation of the plane\cite{harald}.
The convex
polygons of this Voronoi construction are supposed to model 
the grains of the material. 
In this way the structure of the
solid is built on a mesoscopic scale. 
In our simulation these
polygons are the smallest particles interacting 
elastically with each other. 
All the polygons have three continuous
degrees of freedom in two dimensions: the two coordinates 
of the positions of the center of mass and the rotation angle.
 The elastic behavior of the solid is captured in the following way:
The polygons are considered to be rigid bodies.
They are not breakable 
and not deformable.
But they can overlap when they are pressed against each other 
representing to
some extent the local deformation of the grains.
In order to simulate the elastic contact force between touching grains
we introduce a
repulsive  force between the overlapping polygons. This 
force is proportional to the overlapping area divided by a
characteristic length of the interacting polygon pair.
The proportionality factor is the grain bulk Young's modulus $Y$\cite{hjt}.

In order to keep the solid together it is necessary  
to introduce a cohesive force  between 
neighboring polygons. For this purpose we introduce beams,
which were extensively used recently 
in crack growth models\cite{hans,beam}. 
The centers of mass of neighboring
polygons are connected by elastic beams, which  
exert an attractive, restoring force between the grains, 
and can break in order to model
the fragmentation of the solid. The physical properties of the beams,
i.e.. length, section and moment of inertia are determined by the actual
realization of the Voronoi polygons. To describe their elastic
behavior, the beams have a Young's modulus $E$, which is in principal
independent of $Y$, see Ref.~20.

For not too fast deformations  the breaking of a beam is only caused
by stretching and bending.
We  impose a breaking rule of the form of the von Mises plasticity
criterion, which takes into account 
these two breaking modes, and which can reflect the fact 
that the longer and thinner
beams are easier to break. 
The breaking rule contains two parameters 
$t_{\epsilon}$ and $t_{\Theta}$ controlling the relative importance of
the stretching and bending modes.
In the simulations we used the same values of $t_{\epsilon}$ and
$t_{\Theta}$ for all the beams. 
During all the calculations the beams are allowed to break solely under
stretching, which takes into account that it is much harder to break a
solid under compression than under elongation.
The breaking rule is evaluated in each iteration time step and those
beams which fulfill the breaking condition are removed from the
calculation. The simulation stops if there is no beam breaking during
$300$ successive iteration steps. 
Table \ref{table_0} shows the values
of the microscopic
parameters of the model used in the simulations.

\begin{table}[h]
\begin{center}
\vspace{0.4cm}
\caption{ The parameter values used in the simulations.}
\begin{tabular}{l c c c c}
\hline
$Parameter$ & $Symbol$ & $Unit$ & $Value$  \\
\hline
Density & $\rho$ & $g/cm^{3}$ &5  \\
Grain bulk Young's modulus & $Y$ & $dyn/cm^{2}$ & $10^{10}$ \\
Beam Young's modulus & $E$  & $dyn/cm^{2}$ &$5 \cdot 10^9$  \\
Time step & $dt$  & $s$ & $10^{-6}$  \\
The failure elongation of a beam & $t_{\epsilon}$ & $\%$ & $3$ \\
The failure bending of a beam & $t_{\Theta}$ & $degree$ & $4$ \\
\hline
\end{tabular}
\label{table_0}
\end{center}
\end{table}
The calculations were performed on the CM5 of the CNCPST in Paris.
We used the farming method, i.e. the same program runs on a variety of
nodes with different initial setups. In our case 32 nodes were used
with different seeds for the Voronoi generator, i.e. with differently
shaped Voronoi cells.

\section{Results}
With the model outlined above we have already performed several
numerical experiments\cite{ferenc}. Namely, we studied the
fragmentation of a solid
disc caused by an explosion in the middle and the breaking of a
rectangular solid block due to the impact with a projectile.
Emphasis was put on the investigation of the fragment size
distribution. Universal power law behavior was found,
practically independent from the breaking thresholds, see
Ref.~20.  

In the present paper we study the collision of two solid discs of 
equal size.
The disc-shaped granular solid was obtained starting from the
Voronoi tessellation of a square and cutting out a circular window.
This gives rise to a certain roughness of the surface of the
particles. 

The schematic representation of the experimental situation
can be seen in Fig. \ref{fig:impact1}.
In the simulations all the microscopic parameters of the model shown
in Table \ref{table_0} were fixed, only the macroscopic
parameters were varied, i.e. the
initial velocities $\vec{v}_A,\vec{v}_B$ and the radii $R_A, R_B$ of the
colliding particles and the impact
parameter $b$. 
Only the monodisperse case $R_A=R_B$ was considered. The 
velocities of the two particles have the same magnitude and
opposite direction $\vec{v}_A = -\vec{v}_B$. 
The impact parameter $b$ is defined as the distance
between the two centers of mass in the direction perpendicular to the
velocity vectors. $b$ can vary in the interval $[0,R_A+R_B]$.
The range of the initial velocities was chosen to be $12.5-50 m/s$,
and that of the
radii $5-15 cm$.

In the following the results are presented concerning the time
evolution of the fragmenting system, the spatial
distribution of the fragments, the distribution of
the fragment velocities and of the fragment mass. 

\subsection{Fragmentation process}
\label{sec:fragm}
\noindent
In Ref. ~4, based on detailed experimental studies, the
low - velocity impact phenomena of solid spheres were classified into
five categories: $(1)$ elastic rebound, $(2)$ rebound with contact
damage, $(3)$ rebound with longitudinal splits, $(4)$ destruction with
shatter- cone like fragments, and $(5)$ complete destruction.
These categories can be well distinguished by the imparted energy. 
Our simulations cover the
$4th$ and the $5th$ classes varying the initial velocities, the radii and the
impact parameter. Results about the $1st$ case will be
presented in a forthcoming publication.

The collision initiates with the contact of the two bodies. At our
velocity range it can be assumed that the impact proceeds
quasi-statically since the impact velocity is much smaller than the
velocity of the generated shock wave. 
Fig. \ref{fig:impact2} and Fig. \ref{fig:impact3} show representative
examples of the time
evolution of the colliding system at $b/d=0$ (central collision) and
$b/d=0.5$, respectively. Here $d$ denotes the diameter of the
particles. Due to the high compression a strongly damaged region is
formed around the impact site, where practically all
the beams are broken and all the fragments are single polygons. Since
the beam breaking dissipates energy the growth of the damage stops
after some time. 
The shock wave reaching the free boundary gives rise to the expansion of
the particles. This overall expansion initiates crack formation which
results in the final fragmentation of the solids.
The fragments at the anti - impact site of the particles are larger
and they mainly have a shatter cone like shape in agreement with 
experimental observations\cite{exp_colli2}.
Due to the geometry of the system, the fine fragments in the contact 
zone are
confined unless the global expansion sets in. Thus they
undergo many secondary collisions
while the fragments formed  in the outer region can escape without further
interaction. This has an important effect on the velocity distribution of the
debris in the final state.  (See Chapter ~3.2.)

Since the impact velocity is much smaller than the sound speed of the
material, the peak stress produced by the collision around the impact site is
proportional to the normal component of the impact velocity $v_n$ with
respect to the contact surface.
That is why
the final breaking scenarios strongly depend on
the impact parameter. In Fig. \ref{fig:impact2}, in the case of a
central collision the damage is larger, i.e. the shattered
zone is larger and the average fragment size is smaller than in
Fig. \ref{fig:impact3}. The larger $b/d$ (the smaller $v_n$), the less
energy is converted into breaking and the more energy is carried away by
the motion of the fragments. 

In the expanding system the fragments are not isotropically
distributed but the fragment ejection has a preferred direction depending
also on the impact parameter. In Fig
\ref{fig:jet} the jet-like structure of the fragment ejection can be observed,
which means that most of
the fragments are flying in two ``cones'' having a common axis and a
relatively small opening angle.

To determine
the jet-axis we calculated the sphericity $S$ of the velocity
distribution: 
\begin{eqnarray}
  \label{eq:spher}
  S = \min_{\vec{n}} 2 \frac{\sum v_{Ti}^2}{\sum v_i^2},
\end{eqnarray}
where $\vec{v}_i$ is the velocity of the center of mass of fragment
$i$ and $v_{Ti}$ is its transversal component with respect to $\vec{n}$.
The factor $2$ scales the upper limit of $S$ to 1 in the case of the
isotropic velocity distribution. The
jet-axis  $\vec{n}^*$ minimizes the above expression. 
The angle of the jet-axis and that of the contact surface  with
respect to the direction of the initial velocity as a function of
$b/d$ are shown in
Fig. \ref{fig:sphericity}. 

Because of the low impact velocity there is enough time for stress
rearrangement inside the two bodies. Thus the stress can have a tangential
direction to the contact surface giving rise to the jet-like
ejection. (A detailed study of the stress field inside disc shaped
particles due to an impact will be presented in the forthcoming
publication mentioned above.) 
This argument is also
supported by the fact that the long cracks passing through the solids
are either perpendicular to the contact zone or they go radially 
to the surface of the solids.
For the central
and peripheric collisions the jet-axis is practically parallel to the
contact surface (see Fig.\ref{fig:sphericity}).
The two curves differ considerably only at intermediate $b$ values. 

The angular distribution of the fragments
around the direction of the
initial velocity is shown in Fig. \ref{fig:angle}. The concentration
of the debris in a small solid angle can be observed. The position of
the peak of
the distributions practically coincides with the jet-axis.  

\subsection{Scaling of the velocity distribution}
\label{sec:velocity}
\noindent
In order to study the velocity distribution of the fragments we performed two
sets of simulations alternatively fixing
the initial velocity and the
radius of the particles changing the value of the other one. Only
central collisions $b/d=0$ were considered. 
In both cases the distribution of the $x$ and $y$ components of the
velocity of the center of mass of the fragments was evaluated. The $x$
axis was chosen to be parallel
to the initial velocity of the collision.

Fig. \ref{fig:vx_fsize} and Fig. \ref{fig:vx_fvelo} show the results
for fixed $R=15 cm$ varying the initial velocity and for fixed
$v= 50 m/s$ varying the radius of the particles, respectively.
One can observe that the
distributions of both
velocity components always exhibit a Gaussian form.
The zero mean value is a
consequence of momentum conservation. 
In the $x$ direction 
the fragments are slower, i.e. the values of $v_x$ are
much smaller than those of $v_y$. There is a small fraction of the
debris, which has
velocity larger than the initial one, in 
agreement with experimental results\cite{exp_colli3}.   

For fixed system size $R$  one can see 
that the increasing
initial velocity $v$ results in a larger dispersion of the fragment velocities,
increasing the deviation of the
distributions $n(v_x)$, $n(v_y)$ as shown in Fig. \ref{fig:vx_fsize}.   
For fixed initial velocity in Fig. \ref{fig:vx_fvelo}
the deviation of $n(v_x)$ is decreasing with increasing
system size but it remains constant for $n(v_y)$.
The scaling analysis of the velocity distributions is of major interest.
By appropriately rescaling the axes one can collapse the data
obtained at fixed values of the macroscopic parameters 
onto one single curve. We denote the 
velocity components by $v_i$, where
$i=(x,y)$. The data collapse can be obtained using the same form of 
scaling ansatz for both data sets: 
\begin{eqnarray}
  \label{scaling}
  n(v_i)& = & R^{\alpha_i} \phi(v_iR^{\alpha_i}) \hspace{1cm}    \mbox{
    for fixed }  v, \\
  n(v_i)& = & v^{\beta_i} \psi(v_iv^{\beta_i})   \hspace{1.25cm} \mbox{
    for fixed }  R, \label{beta}
\end{eqnarray}
where $\phi, \psi$ are scaling function and $\alpha_i, \beta_i$
are the scaling
exponents belonging to the velocity component $v_i$. 
In principle, one could introduce two different
exponents for the macroscopic variables
within Eqs. (\ref{scaling}, \ref{beta}) but in our case they turned 
to be equal within the error bars. The values of $\alpha_i$ and $\beta_i$
are presented in Table \ref{tab:alfa}. 
\begin{table}[htbp]
\begin{minipage}[h]{\textwidth}
\begin{center}
\caption{ The values of the scaling exponents $\alpha_i, \beta_i$ for 
  $n(v_x)$, $n(v_y)$.}
\begin{tabular}{l c c c}
\hline
& $\alpha_i$ & $\beta_i$ \\ 
\hline
 $n(v_x)$  &$-0.7 \pm 0.05$  & $0.35 \pm 0.04$ \\ 
 $n(v_y)$  &$ 0 \pm 0.02  $  & $0.85 \pm 0.05$ \\ 
\hline
\end{tabular}
\label{tab:alfa}
\end{center}
\end{minipage}
\end{table}
Note, that in both cases the
scaling exponents of the two velocity components $\alpha_x, \alpha_y$ and
$\beta_x, \beta_y$ are significantly different. 

The data collapse obtained using Eqs. (\ref{scaling}, \ref{beta}) is
illustrated
by Figs. \ref{fig:vx_fsize_sc}, \ref{fig:gauss}. The quality of the
collapse is satisfactory, there are fluctuation only for the smallest
initial velocity in Fig. \ref{fig:vx_fsize_sc} and for the smallest
radius in Fig. \ref{fig:gauss} when the number of fragments is
not large enough. The scaling functions $\phi, \psi$ are the same for
the two velocity components within the accuracy of the calculation.
The scaling structure found implies that the distributions $n(v_i)$ are
Gaussians, the standard deviation $\sigma_i$ of which has a power law
dependence on the macroscopic parameters, i.e.: 
\begin{eqnarray}
  \label{gaus}
  n(v_i)& = & \frac{1}{\sqrt{2\pi} \sigma_i}
  \exp{\left[ -\frac{1}{2}
      \left(\frac{v_i}{\sigma_i}\right)^2\right]}, \\ 
  \sigma_i(R,v) & \sim & R^{\alpha_i}v^{\beta_i}.   \label{sigma}      
\end{eqnarray}
Fig. \ref{fig:gauss} shows a representative example of the
Gaussian fit too. Eq. (\ref{gaus}) was fitted to the scaling 
function $\phi$.  

The shape of the velocity distribution of the fragments is mainly
determined by the
stress distribution in the bodies just before the breaking of the
beams.  
When, due to the geometry of the system,
energetic fragments are confined during some time, secondary
collisions of the
products can also have a considerable contribution. In our case this
effect can be responsible for the Gaussian shape.

The deviations of the two velocity components
$\sigma_x$ and $\sigma_y$ can be considered as the linear
extensions of the fragment jet in the velocity space in the $x$ and
$y$ directions. 
Their ratio $s$ is a
characteristic quantity of the jet shape.
(Note, that here we considered solely central
collisions. The above argument can be generalized to non-central
collisions choosing the $x$ direction perpendicular to the jet-axis.)
Eq. (\ref{sigma}) yields
the dependence of $s$ on the macroscopic parameters:
\begin{eqnarray}
  \label{sh}
   s(R,v)=\frac{\sigma_y}{\sigma_x} \sim
   R^{\alpha_y-\alpha_x}v^{\beta_y-\beta_x} 
   = R^{\gamma}v^{\delta}
\end{eqnarray}
The value of the new exponents characterizing the jet-shape are $\gamma =
0.7 \pm 0.05$ and  $\delta = 0.5 \pm 0.05$.

 It is generally believed that once
a solid, e.g. asteroid suffers catastrophic destruction the pieces fly
apart from each other in all directions. However this scheme is not 
necessarily true. 
From our treatment it follows that we have anisotropic clustering
of fragments, i.e. most of
the fragment velocities are much smaller than the impact velocity and
the particles fly in rather collimated jets, the shape of which is
described by $\gamma$ and $\delta$.
This can have important consequences for the later
evolution of the fragmented system. For instance,
in the case
of the collision of asteroids fragments could recombine
due to mutual gravitation forming a cloud-like object. 
\vspace*{1cm}

\subsection{The fragment mass distribution}
\noindent
\label{sec:mass}
Beside the fragment velocities, the mass distribution of the debris
also has practical importance.
Fig. \ref{fig:impact2} and Fig. \ref{fig:impact3} show
that the contact zone around the
impact site
gives the main contribution to the small fragments while
the larger ones are dominated by the anti-impact site. This detachment
effect 
becomes more pronounced for smaller collision velocities when the
type of the impact passes from class $(5)$ to $(4)$ (see Chapter
~3.1).

 The fragment mass histograms $F(m)$ are presented in
Figs. \ref{fig:mass_fsize}, \ref{fig:mass_fvelo} 
for fixed
system size and for fixed initial velocity, 
respectively.  Here $F(m)$ denotes the number of fragments with
mass $m$ divided by
the total number of fragments.
In order to obtain the correct shape of the distributions at
the coarse products as well as at the finer ones logarithmic bining
was used, i.e. the bining is equidistant on logarithmic scale.
The histograms have two cutoffs. The lower one is due to 
the existence of single unbreakable polygons (smallest
fragments) and the upper one is given by the finite size of the
system (largest fragment). 

In Fig. \ref{fig:mass_fsize} the histogram
belonging to the smallest collision velocity has
two well distinguished local maxima, one for the fine fragments
(single polygons and 
pairs) and another one for the large pieces, which are comparable
to the size of the colliding particles. In between, for the
``Intermediate Mass Fragments'' (IMF) $F(m)$ shows
power law behavior, i.e. we seem to have:
\begin{eqnarray}
  \label{power}
  F(m) \sim m^{-\mu}.
\end{eqnarray}
 The effective exponent $\mu$ was
obtained from the estimated slope of the curve, $\mu = 2.1 \pm 0.05$
for $v=12.5 m/s$.  At low impact velocity this shape of distribution
is characteristic for light-fragment
ejection from a heavy system\cite{frag_book}.
As the initial velocity increases the peak of the heavy fragments 
gradually disappears giving rise to an increase of the contribution of IMFs, 
while the fraction of the shattered products hardly changes. 
In the IMF region the slope of $F(m)$ depends slightly on the
impact velocity, i.e. the distributions become less steep
with increasing $v$.
The straight line in Figs. \ref{fig:mass_fsize}, \ref{fig:mass_fvelo} 
shows the power law fitted at our highest velocity $v=50 m/s$ and for
$R=15 cm$ with exponent 
$\mu =1.75 \pm
0.05$.

For fixed initial velocity  the histograms belonging to different system
sizes are characterized practically by the same exponent $\mu = 1.75
\pm 0.05$ as shown in Fig. \ref{fig:mass_fvelo}.
Only in the case of the smallest system, which suffers
catastrophic destruction shattering the particles completely, the
exponent is larger.

It is important to note that in nuclear fragmentation experiments of
gold projectile with several targets, the
charge distribution of the fragments shows a similar dependence on the
deposited energy of the collision\cite{aladin}. 

Beside the size distribution of the debris there is also
interest in the energy required to achieve a certain size
reduction. It was revealed in collision experiments\cite{exp_colli2}
that the mass of
the largest fragment $M_{max}$ normalized by the total mass $M_{tot}$
follows a power law as
a function of the specific energy, i.e. 
the ratio of the imparted energy $E$ and the total mass of the system. 
The exponent characterizing the size reduction seems to be independent
of the type of the material but it is sensitive to the shape of the
colliding particles. For spherical bodies it was found to be around
$0.7$, while for cubic systems around $1.5$ (see
Ref.~4).  
Note, that in Fig. \ref{fig:mass_fvelo} the fixed initial velocity
implies that the specific energy is also constant. This results in
more or less 
the same value of $M_{max}/M_{tot}$ for the different curves. In 
Fig. \ref{fig:mass_fsize} for fixed system size the specific energy
varies with the initial
velocity resulting change of $M_{max}/M_{tot}$.

In order to obtain more information about the size reduction we performed a
set of simulations on the $32$ partition of the CM5 with fixed $R= 15
cm$ particle radius changing the initial velocity within the interval
$12.5-50 m/s$ in 32 steps.
In Fig. \ref{fig:degrad} $M_{max}/M_{tot}$ is plotted against the
specific energy $E/M_{tot}$ on double logarithmic scale. Although the
data are rather scattered
a power law seems to be a reasonable fit with exponent $0.68$ in 
agreement with the experimental results. 

\section{Conclusion}
\noindent
We have studied the low velocity impact phenomena of two solid discs
of the same size using our cell model\cite{ferenc}. We focused our
attention on the spatial distribution of the debris and on the
analysis of the fragment velocities and fragment mass.

Anisotropic clustering of fragments was revealed, which manifests in
the jet structure of the fragment ejection. Due to the anomalous
scaling of the distributions of the velocity components the jet-shape
can be characterized by two independent exponents.  
The mass distribution of the intermediate mass fragments shows a power
law behavior, whose exponent slightly decreases with increasing imparted
energy. 
We have noted that the charge
distributions obtained  in nuclear multifragmentation experiments
show a similar dependence on the deposited energy of the
collision.
This is a manifestation of the 
independence of the global quantities of the fragmenting system from the
microscopic details of the mechanism of the individual breaking.

The mass of the largest fragment normalized by the total mass also
follows a power law as a function of the imparted energy density
with exponent close to the experimental observations.

Still our study makes a certain number of
technical simplifications which might be important for a full
quantitative grasp of fragmentation phenomena. Most important seems to
us the restriction to two dimensions, which should be overcome in
future investigations. The existence of elementary, non-breakable
polygons restricts fragmentation on lower scales and hinders us from
observing the formation of a powder of a shattering transition\cite{redner}. 

Experiments showed\cite{exp_colli2} that the relative size 
of the colliding bodies is an important parameter describing the low
velocity impact phenomena if the mechanical properties of the bodies
are the same. In the future our studies should be extended to this
polydisperse case too.

\section*{Acknowledgment}
\noindent
We are grateful to CNCPST in Paris for the computer time on the
CM5. F. Kun acknowledges the financial support 
of the Hungarian Academy of Sciences and the EMSPS.

\begin{figure}[!h]
\vspace{0.3cm}
\begin{center}
\epsfig{bbllx=72,bblly=286,bburx=407,bbury=529,file=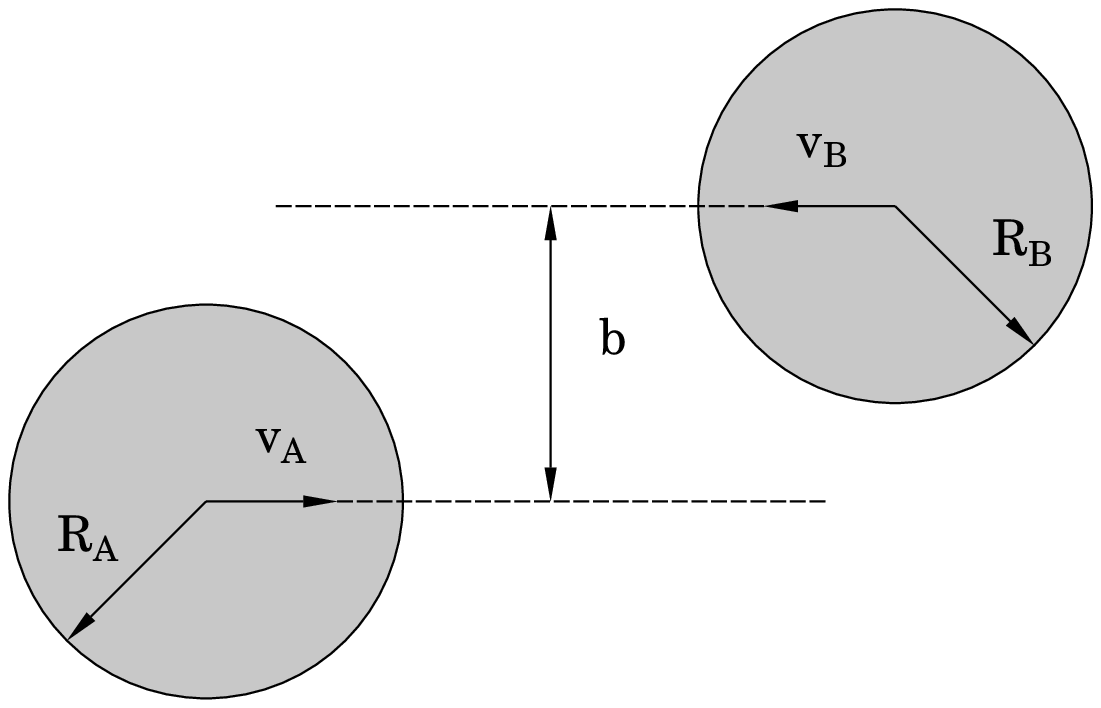,width=6cm}
\caption{The schematic representation of the collision experiment.
 In all the simulations $R_A = R_B$ and $\vec{v}_A=-\vec{v}_B$ were
 chosen.}
\label{fig:impact1}
\end{center} 
\end{figure}

\begin{figure}[!h]
\vspace{0.3cm}
\begin{center}
\epsfig{bbllx=175,bblly=120,bburx=500,bbury=780,file=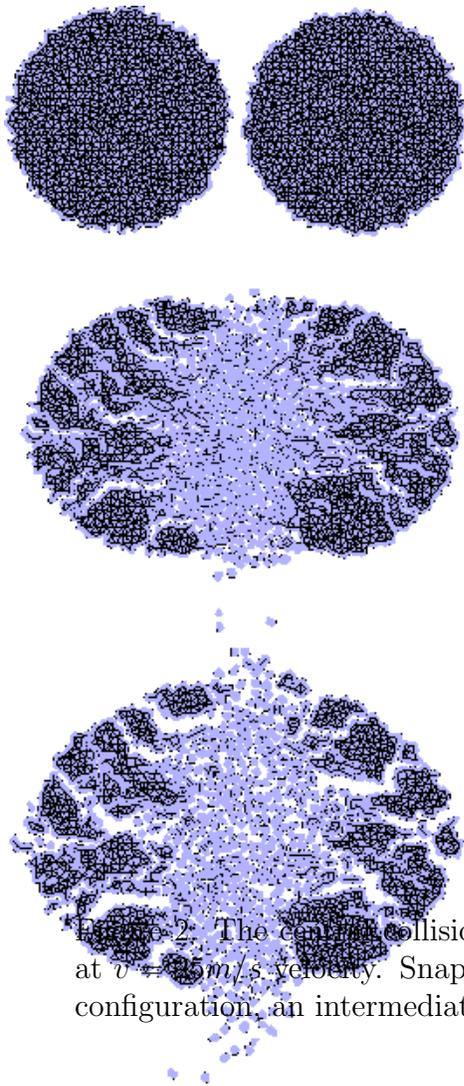,width=7.5cm}
\caption{The central collision ($b=0$) of two discs of equal size
 $R=15 cm$ at $v = 25 m/s$ velocity.
 Snapshots of the evolving system present the initial configuration,
 an intermediate state and the final breaking scenario.}   
\label{fig:impact2}
\end{center} 
\end{figure}

\begin{figure}[!h]
\vspace{0.3cm}
\begin{center}
\epsfig{bbllx=175,bblly=120,bburx=500,bbury=860,file=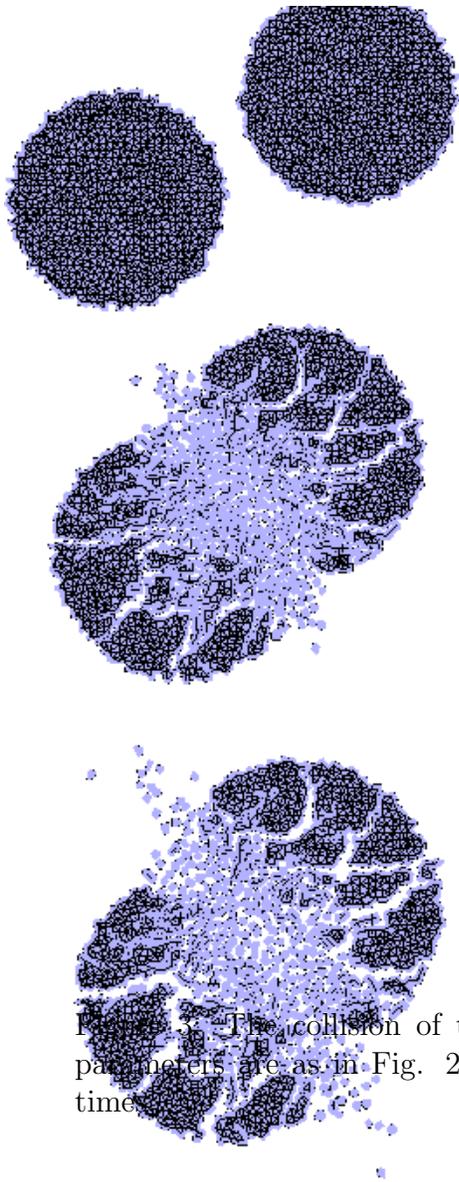,width=7.5cm}
\caption{The collision of two discs of equal size at $b/d=0.5$.
  All the parameters are as in Fig. \ref{fig:impact2} and the
  snapshots are also taken at the same times.}
\label{fig:impact3}
\end{center} 
\end{figure}

\begin{figure}[!h]
\vspace{0.3cm}
\begin{center}
\epsfig{bbllx=53,bblly=342,bburx=501,bbury=690,file=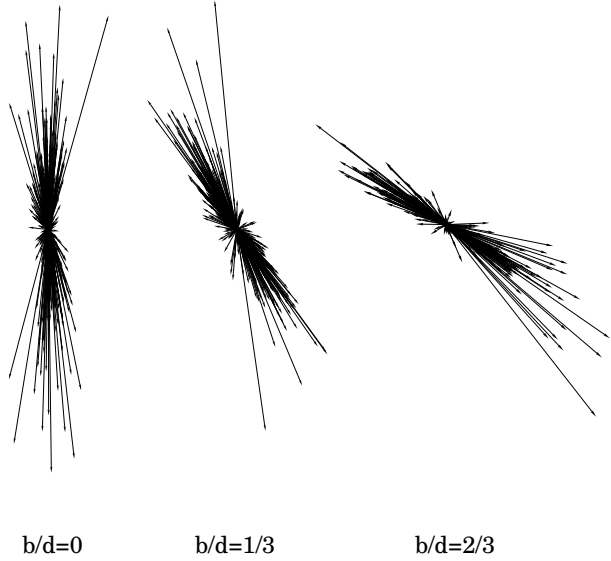,width=10cm}
\caption{The velocity vectors of the center of mass of the fragments
 for three different impact parameters. One can observe the jet
 structure of the fragment ejection.
 All the parameters are as in
 Fig. \ref{fig:impact2}.} 
\label{fig:jet}
\end{center} 
\end{figure}

\begin{figure}[!h]
\vspace{0.3cm}
\begin{center}
\epsfig{bbllx=80,bblly=325,bburx=445,bbury=656,file=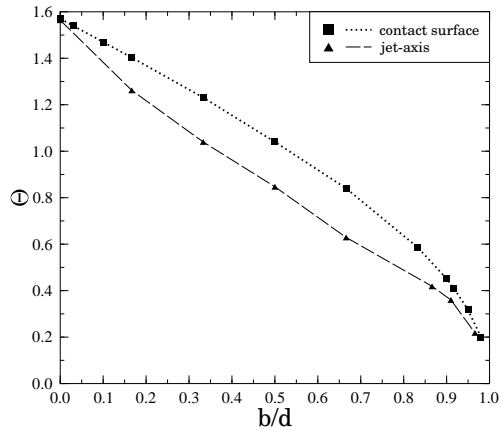,width=7cm}
\caption{The angle of the jet-axis and that of the contact surface
  with respect to the direction of
  the initial velocity as a function of $b/d$.}
\label{fig:sphericity}
\end{center} 
\end{figure}

\begin{figure}[!h]
\vspace{0.3cm}
\begin{center}
\epsfig{bbllx=80,bblly=325,bburx=445,bbury=656,file=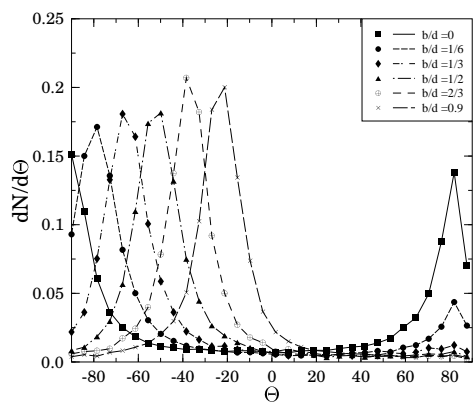,width=7cm}
\caption{The angular distribution of the fragments around the
  direction of the initial velocities for different impact parameters.}
\label{fig:angle}
\end{center} 
\end{figure}

\begin{figure}[!h]
\vspace{0.3cm}
\begin{center}
\epsfig{bbllx=122,bblly=8,bburx=493,bbury=650,file=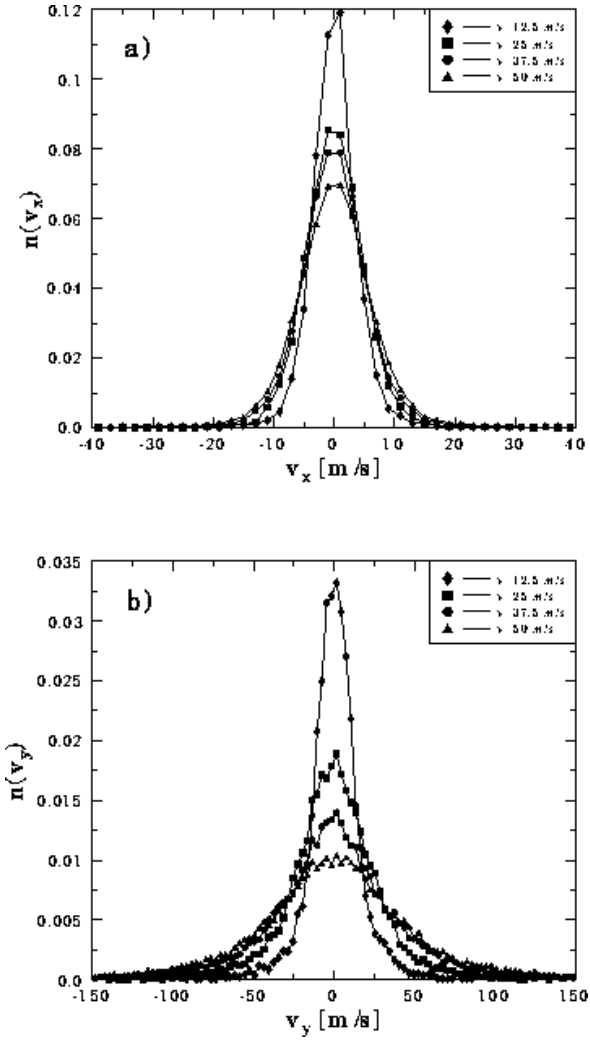,width=8cm}
\caption{The distribution of the $x$ and $y$ components of the velocity of the
  fragments with fixed system size $R = 15 cm$ varying the
  initial velocity $v$.}
\label{fig:vx_fsize}
\end{center} 
\end{figure}

\begin{figure}[!h]
\vspace{0.3cm}
\begin{center}
\epsfig{bbllx=122,bblly=8,bburx=493,bbury=650,file=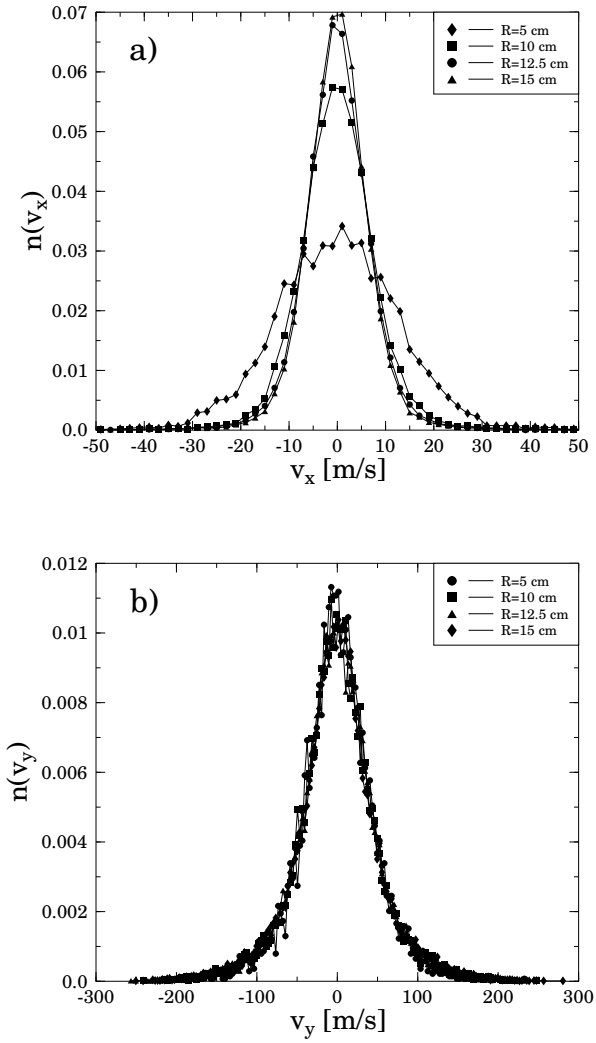,width=8cm}
\caption{The distribution of the $x$ and $y$ components of the velocity of the
  fragments with fixed initial velocity $v = 50 m/s$ varying the
  radius $R$ of the particles.}  
\label{fig:vx_fvelo}
\end{center} 
\end{figure}

\begin{figure}[!h]
\vspace{0.3cm}
\begin{center}
\epsfig{bbllx=122,bblly=0,bburx=493,bbury=650,file=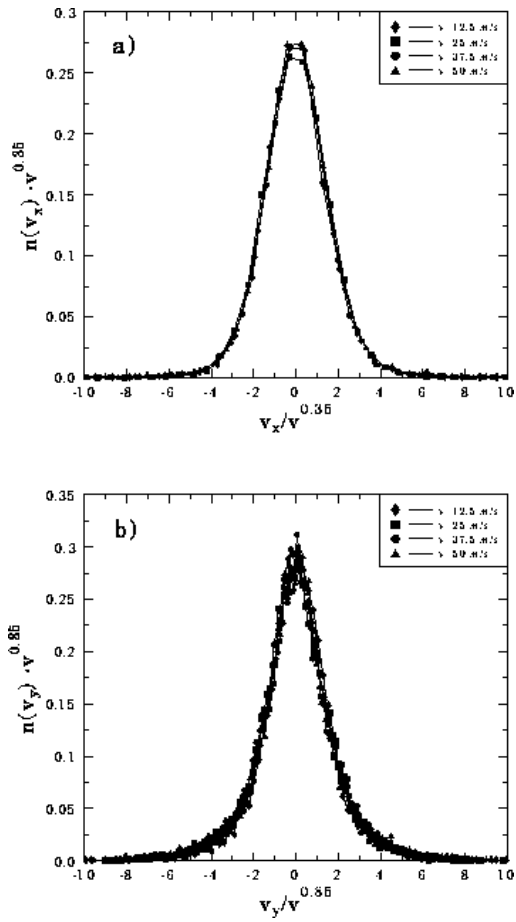,width=7cm}
\caption{Scaling of the velocity distributions for fixed system
  size $R=15 cm$ varying the initial velocity $v$.}
 \label{fig:vx_fsize_sc}
\end{center} 
\end{figure}

\begin{figure}[!h]
\vspace{0.3cm}
\begin{center}
\epsfig{bbllx=122,bblly=300,bburx=493,bbury=650,file=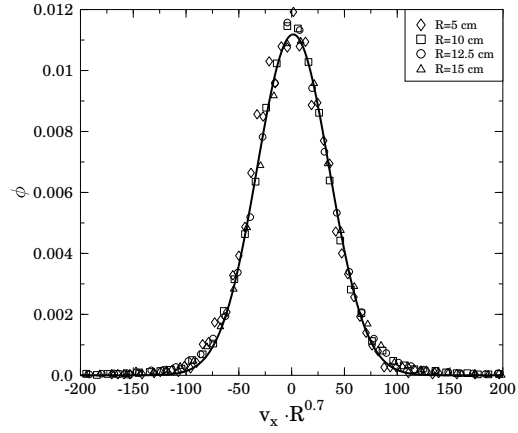,
 width=7cm}
\caption{Scaling of the velocity distributions for fixed initial
 velocity $v=50 m/s$ varying the radius $R$  of the particles. The
 solid line shows  the Gaussian fit according to
  Eq. (\ref{gaus}) for the scaling
  function $\phi$.}
 \label{fig:gauss}
\end{center} 
\end{figure}

\begin{figure}[!h]
\vspace{0.3cm}
\begin{center}
\epsfig{bbllx=122,bblly=300,bburx=493,bbury=630,file=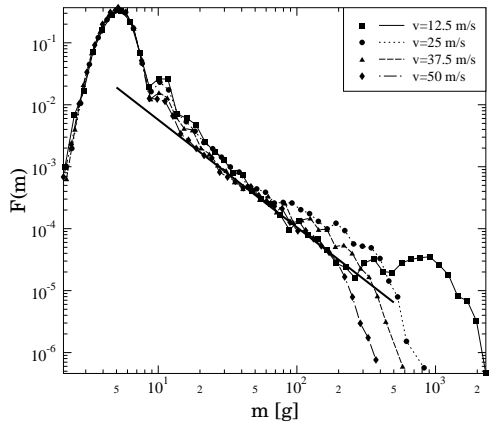,
 width=7cm}
\caption{The fragment mass histograms for fixed system size $R=15 cm$
  varying the initial velocity $v$. The straight line shows the
  power law fitted to the curve belonging to $v=50 m/s$.}
   \label{fig:mass_fsize}
\end{center} 
\end{figure}

\begin{figure}[!h]
\vspace{0.3cm}
\begin{center}
\epsfig{bbllx=122,bblly=300,bburx=493,bbury=630,file=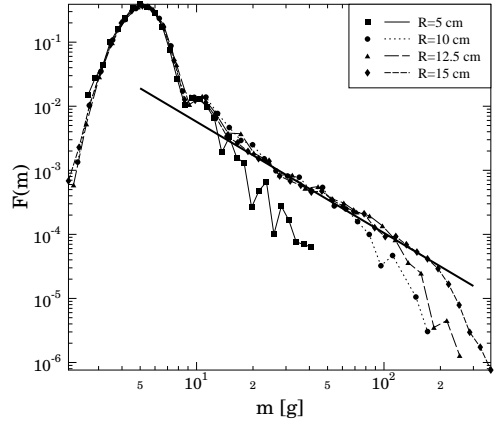,
 width=7cm}
\caption{The fragment mass histograms for fixed initial velocity $v=50
  m/s$, varying the system size $R$. The straight line is the same as
  in Fig. \ref{fig:mass_fsize}.} 
 \label{fig:mass_fvelo}
\end{center} 
\end{figure}

\begin{figure}[!h]
\vspace{0.3cm}
\begin{center}
\epsfig{bbllx=122,bblly=300,bburx=493,bbury=630,file=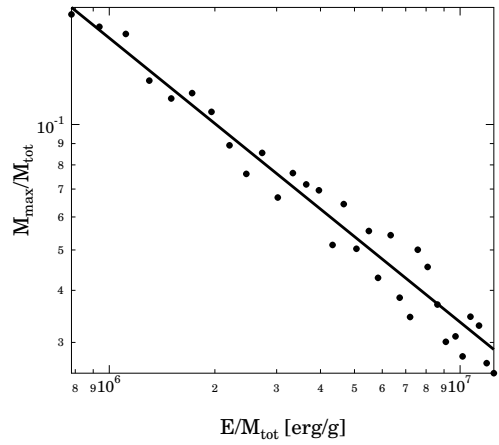,
 width=7cm}
\caption{The mass of the largest fragment normalized by the total mass
  as a function of the specific energy.}
 \label{fig:degrad}

\end{center} 
\end{figure}

\end{document}